\documentclass[reprint,amsmath,amssymb,aps]{revtex4-1}
\usepackage{lipsum}
\usepackage{graphicx}
\usepackage{dcolumn}
\usepackage{bm}
\usepackage{natbib}
\newcommand{\sgn}{\rm sign}
\begin{document}
\preprint{APS/123-QED}
\title{Interaction induced local moments in parallel quantum dots \\within the functional renormalization group approach}
\author{V. S. Protsenko}
\author{A. A. Katanin}
\affiliation{%
M. N. Mikheev Institute of Metal Physics, 620990 Ekaterinburg Russia\\
Ural Federal University, 620002 Ekaterinburg, Russia
}
\date{\today}

\begin{abstract}
We propose a version of functional  renormalization-group (fRG) approach, which is, due to including Litim-type cutoff and switching off (or reducing) the magnetic field during fRG flow, capable describing singular Fermi liquid (SFL) phase, formed due to presence of local moments in quantum dot structures. The proposed scheme allows to describe the first-order quantum phase transition from "singular" to the "regular" paramagnetic phase with applied gate voltage to parallel quantum dots, symmetrically coupled to leads, and shows sizable spin splitting of electronic states in the SFL phase in the limit of vanishing magnetic field $H\rightarrow 0$; the calculated conductance shows good agreement with the results of the numerical renormalization group. Using the proposed fRG approach with the counterterm, we also show that for asymmetric coupling of the leads to the dots the SFL behavior similar to that for the symmetric case persists, but with occupation numbers, effective energy levels and conductance changing continuously through the quantum phase transition into SFL phase. 
%For asymmetric coupling we find Fano-like resonance of the conductance near the transition to the SFL state.  
\end{abstract}
\maketitle
\section{Introduction} 
During last decades electronic and transport properties of quantum dots have attracted considerable attention due to possibility of realizing non-trivial quantum phenomena and  promising applications in different nanodevices \cite{Kou0,Song}, in particular quantum computation systems \cite{Awschalom,Bennett,Loss,Engel,Hanson,Petta}. 
The electron-electron interaction may have significant effect on electronic properties and transport in quantum dot systems and is responsible for many interesting physical phenomena. 

While relatively strong Coulomb interaction leads to Kondo effect \cite{Hewson_1997,Goldhaber-Gordon,Cronenwett2,van_der_Wiel_2}, for some special geometries of quantum dot systems  
even relatively weak interaction plays important role. 
This concerns in particular a system of parallel quantum dots, connected symmetrically to common leads,
where the formation of the so called singular Fermi liquid (SFL) state \cite{Zitko_1,Zitko_2}, which possesses a local moment, decoupled from the leads, and 
almost unitary conductance, is possible near half filling. 

To date, a wide variety of analytical and numerical methods have been developed to describe the effect of the Coulomb interaction in quantum dot structures. The  numerical renormalization group (NRG) method provides a strightforward approach to describe spectral functions and conductivity of quantum dots in the presence of an interaction \cite{Oguri_2,Oguri_3,Oguri_4,Lopez}. However, being very successful, this method requires significant  computational effort, which grows exponentially with the number of interacting degrees of freedom.  This circumstance does not allow one to directly apply these methods to fairly large systems. 

The recently prposed nanoscopic dynamic vertex approximation (nano-D$\Gamma$A) \cite{nanoDGA1,nanoDGA2}, allows one to treat effects of interaction in nanostructures, and has a potential of describing non-local correlations beyond nano-DMFT \cite{nanoDMFT1,nanoDMFT2,nanoDMFT3,nanoDMFT4,nanoDMFT5}. 
In the most complete, parquete, formulation nano-D$\Gamma$A approach however also requires substantial computational resources.

In this regard, the recently proposed functional renormalization-group (fRG) approach \cite{Wetterich,Salmhofer_1,Metzner} 
is very promising because it allows to reformulate a many particle problem in terms of coupled differential equations for irreducible vertex functions, which, after several approximations, can be easily integrated even for complex systems. This method has already been successfully applied to systems consisting of a small number of quantum dots arranged in different geometries \cite{Andergassen,Karrasch,Meden_1,Meden_2}. In these studies it was shown that the results obtained with the 
sharp frequency cutoff 
are in good agreement with the Bethe ansatz and NRG data \cite{Karrasch}.
At the same time,
the fRG approach with the sharp cutoff scheme is not applicable to describe the SFL state, since, e.g., for parallel quantum dots it yields an artificially low conductance at low magnetic fields for gate voltages $V_{g}$ less than some critical value $\left|V_{g}\right|\lesssim V_{g}^c$ \cite{Karrasch}, which contradicts the NRG results \cite{Zitko_1}. \par
To describe physical properties in the local moment SFL state, we propose in the present paper an fRG scheme with continously switching off (or decreasing) the external magnetic field with decreasing cutoff. This scheme is similar to a counterterm technique, used earlier to treat the first order quantum phase transitions in lattice fermionic systems~\cite{Gersch_1,Gersch_2} within fRG approach. This technique was not however applied to description of quantum dot structures. We show that within this approach it is possible to 
describe correctly the dependence of the conductance on the magnetic field  and the gate voltage
in a good agreement with NRG data. We also found that the influence of an asymmetrical dot-lead coupling can be straightforwardly investigated using the same formalism. To demonstrate this, as an example, we consider the particular case of a parallel quantum dot system with different coupling of the leads to the dots. An especially interesting observation is the continuous behavior of the calculated observables at the quantum phase transition to the SFL state in this case, leading to the resonance form of the conductance curve in the vicinity of the phase transition point.

The plan of the paper is the following. In Sect. II we formulate the model of parallel quantum dots with arbitrary couplings to the leads. In Sect. III we introduce the fRG approach with the counterterm, compare its results to the standard fRG approach and numerical renormalization group data. In Sect. IV we analyze presence of the local moments and the origin of spin splitting of electronic spectrum in the SFL regime, as well as the conductance of quantum dots, symmetrically and asymmetrically coupled to the leads. Finally, in Sect. V we present the conclusions. 

\section{Model} 
\begin{figure}[t] 
\centering
\vspace{-0.1cm}
\includegraphics[width=0.45\linewidth]{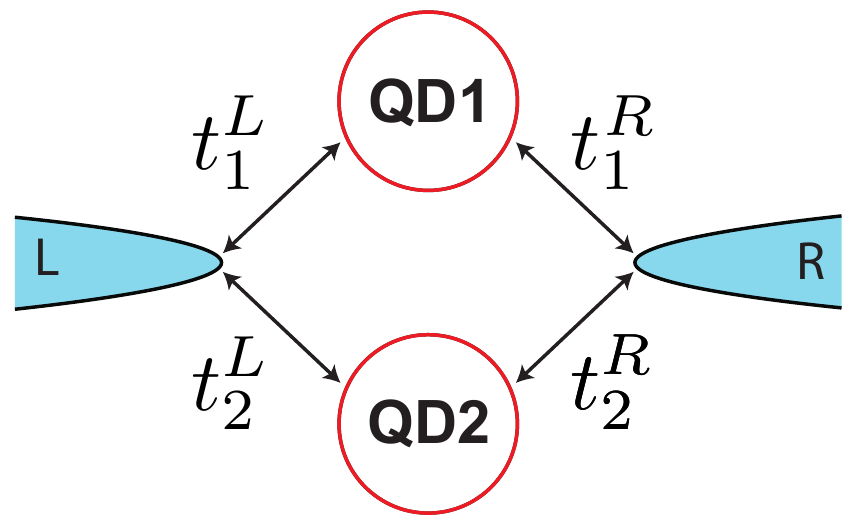}
\hspace{0.2cm}  \includegraphics[width=0.45\linewidth]{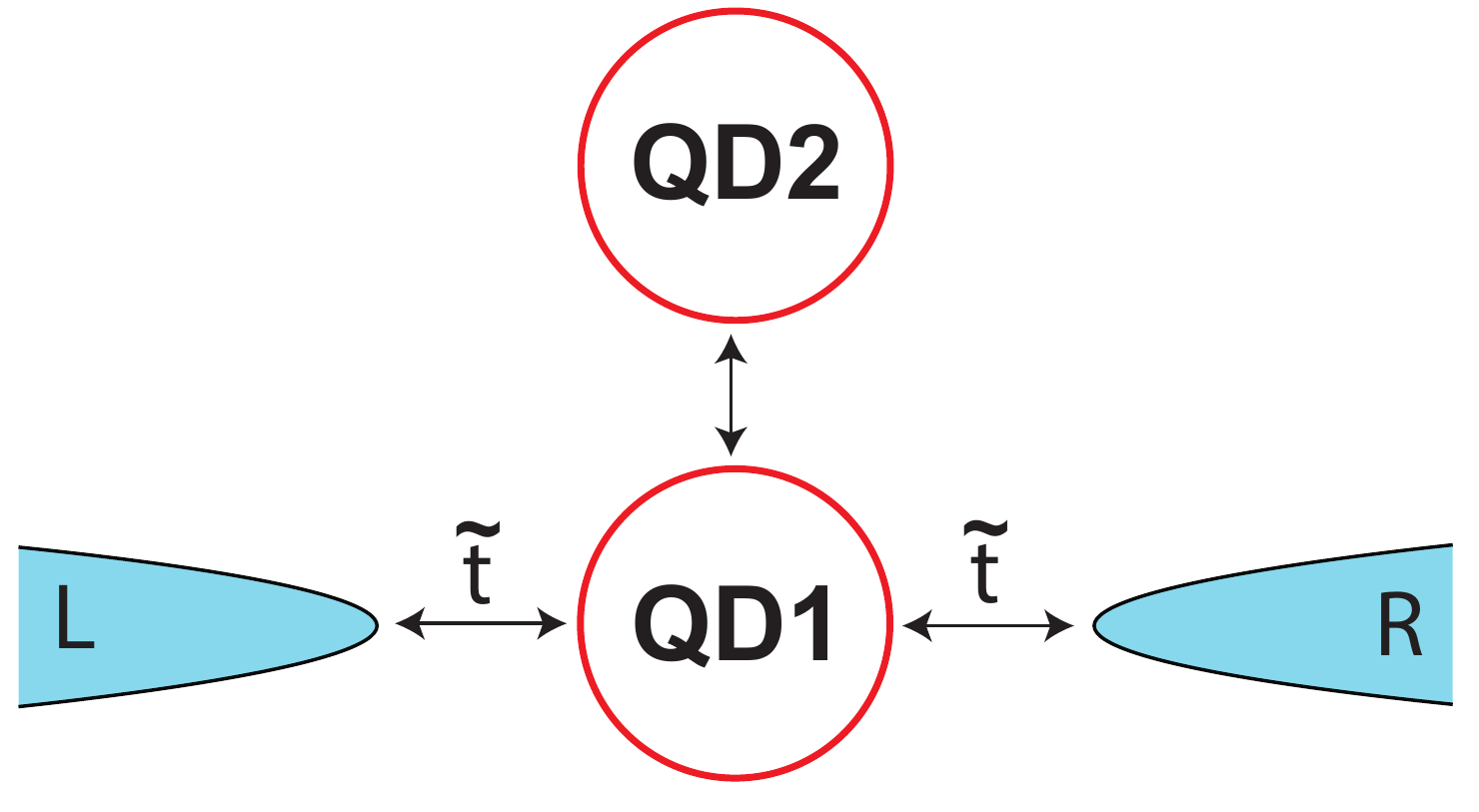}
\caption{Left: Schematic representation of parallel double quantum dots (DQ1 and DQ2) connected to two left (L) and right (R) leads by equal tunneling matrix elements $t^\alpha_i$. Right: Schematic representation of a configuration of parallel quantum dots, symmetrically coupled to the leads, obtained as a result of the transformation~(\ref{even-odd orbitals}). The vertical up--down arrow indicates the presence of two-particle interactions between quantum dots (see Eq.~\ref{H_e_o} in the Section ~\ref{Symm_coupl} of the paper).}
\vspace{-0.1cm}
\label{Side_coupled_dots}
\end{figure}

The parallel quantum dots connected to two conducting leads (see  Fig.~\ref{Side_coupled_dots}) can be modelled by the Hamiltonian~\cite{Karrasch}
\begin{equation}
\mathcal{H}=\mathcal{H_{\rm dot}}+\mathcal{H}_{leads}+\mathcal{H}_{coup}, 
\end{equation}
where
\begin{equation}
 \mathcal{H_{\rm dot}}=\sum_{j \sigma}\left[\epsilon^{0}_{\sigma}n_{j,\sigma}+\dfrac{U}{2}\left(n_{j,\sigma}-\dfrac{1}{2}\right)\left(n_{j,\bar{\sigma}}-\dfrac{1}{2}\right)\right]
 \label{H}
\end{equation} 
describes quantum dots with equal on--site Coulomb repulsion $U$ and single energy level $\epsilon^{0}_{\sigma}=V_{g}-\sigma H/2$ ($\sigma=+1/-1$ for the spin up/down electrons), controlled by the gate voltage $V_{g}$ and magnetic field $H$, $d^{\dagger}_{j,\sigma} (d_{j,\sigma})$ -- denote creation (annihilation) operators for an electron with spin $\sigma$ localized on the $j$-th quantum dot and $n_{j,\sigma}=d^{\dagger}_{j,\sigma}d_{j,\sigma}$. The next term, $\mathcal{H}_{leads}$, takes into account the two equivalent non-interacting semi--infinite leads, which are modeled by the following tight-binding Hamiltonian
\begin{equation}
\mathcal{H}_{leads}=-\tau\sum_{\alpha=L,R}\sum_{j=0}^{\infty}\sum_{\sigma}( c^{\dagger}_{\alpha,j+1,\sigma}c_{\alpha,j,\sigma}+\text{H.c.}),
\end{equation}
here $c^{\dagger}_{\alpha,j,\sigma}\textbf{ }( c_{\alpha,j,\sigma})$ are the leads creation (annihilation) operators  for an electron with spin direction $\sigma$ on the $j$ lattice site of the left $\alpha=L$ or right $\alpha=R$ lead and $\tau$ is the hopping matrix element in the leads.
Finally, the last term $\mathcal{H}_{coup}$ introduces coupling between each quantum dot $i$ and leads  $\alpha$, and is given by
\begin{equation}
\mathcal{H}_{coup}=-\sum_{\alpha=L,R}\sum_{i,\sigma}(t^{\alpha}_{i}c^{\dagger}_{\alpha,0,\sigma}d_{i,\sigma}+\text{H.c.}),
%\mathcal{H}_{coupl}=-t\sum_{\alpha=L,R}\sum_{i,\sigma}(c^{\dagger}_{\alpha,0,\sigma}d_{j,\sigma}+\text{H.c.}),
\end{equation}
where $t^\alpha_i$ are the corresponding hopping parameters ($i=1,2, \alpha=L,R$).\par %which causes the dots--leads hybridization strength $\Gamma^\alpha_{i}=\pi \rho_{\text{lead}}(0) |t_{i}^\alpha|^{2}$, where $\rho_{\text{lead}}$ is the local density of states at the last site of lead.\par
 For the symmetric case $t_i^\alpha=t$ one can apply canonical transformation to the even $d_{e,\sigma}$ and the odd $d_{o,\sigma}$ orbitals 
\begin{equation}
d_{e(o),\sigma}=(d_{1,\sigma}\pm d_{2,\sigma})/\sqrt{2},
\label{even-odd orbitals}
\end{equation}
such that the coupling part $\mathcal{H}_{coup}$ takes the form
\begin{equation}
\mathcal{H}_{coup}=-\tilde{t}\sum_{\alpha=L,R}\sum_{\sigma}(c^{\dagger}_{\alpha,0,\sigma}d_{e,\sigma}+\text{H.c.}),
\end{equation}
therefore only the even orbitals are directly connected to the leads by the hopping amplitude $\tilde{t}=\sqrt{2}t$ and the parallel double dot system considered in the present study can be equivalently mapped onto the system is shown on the right side of the Fig.~\ref{Side_coupled_dots}.\par
After projection of the leads and taking the wide--band limit (see Refs.~\cite{Karrasch, Karrasch_thesis,Enss_thesis}), the inverse of the noninteracting propagator for electrons with spin $\sigma$ represents the $2\times 2$ matrix in the dot-space, which has the following structure
\begin{multline}
\left[\mathcal{G}_{0,\sigma}^{-1}(i\omega)\right]_{i,j}=\left[i\omega-\epsilon^{0}_{\sigma}+i\left(\Gamma_{i}^{L}+\Gamma_{i}^{R}\right)\sgn(\omega)\right]\delta_{ij}\\+i\sum_{\alpha=L,R}\sqrt{\Gamma_{i}^{\alpha}\Gamma_{j}^{\alpha}}\sgn(\omega)(1-\delta_{ij}),
\label{green_function}
\end{multline}
with the energy independent hybridization strength $\Gamma_i^\alpha=\pi|t_i^\alpha|^{2}\rho_{\text{lead}}(0)$, where $\rho_{\text{lead}}$ is the local density of states at the last site of the left or right lead (the leads are equivalent).\par
\section{fRG approach} 
\subsection{Formulation of the method}
To describe the correlation effects in quantum dots at zero temperature $T=0$, we, following to Ref. \cite{Karrasch}, project out the leads, introduce some cutoff of the bare single-particle Green function of the dots $\mathcal{G}_{0}\rightarrow\mathcal{G}_{0}^{\Lambda}$, and apply  the one-particle irreducible fRG scheme \cite{Metzner,Salmhofer_1}, yielding an infinite hierarchy of differential flow equations 
for the self-energy $\Sigma$ and the $n$--particle vertices $\Gamma^{(2n)}$, $n\geq2$.  
Truncating the fRG flow equations by neglecting of the flow of the vertex functions with $n\geq3$ leads to a closed system of the flow equations for the $\Sigma$ and the two-particle vertex $\Gamma^{(4)}\equiv\Gamma$, which has the standard form \cite{Andergassen,Karrasch} 
\begin{equation}
\partial_{\Lambda}\Sigma^{\Lambda}_{\alpha^{'}\alpha}=-\int{\dfrac{d\omega}{2\pi} e^{i\omega 0^{+}}\mathcal{S}^{\Lambda}_{\beta\beta^{'}}}\left(i\omega\right)\Gamma^{\Lambda}_{\alpha^{'}\beta^{'}\alpha\beta},
\end{equation}
and  
\begin{align}
\partial_{\Lambda}\Gamma^{\Lambda}_{\alpha^{'}\beta^{'}\alpha\beta}&=\int{\dfrac{d\omega}{2\pi}\mathcal{S}^{\Lambda}_{\gamma\gamma^{'}}\left(i\omega\right)\mathcal{G}^{\Lambda}_{\delta\delta^{'}}\left(-i\omega\right)}
\Gamma^{\Lambda}_{\alpha^{'}\beta^{'}\delta\gamma}\Gamma^{\Lambda}_{\delta^{'}\gamma^{'}\alpha\beta}\nonumber\\&-\int{\dfrac{d\omega}{2\pi}\mathcal{S}^{\Lambda}_{\gamma\gamma^{'}}\left(i\omega\right)\mathcal{G}^{\Lambda}_{\delta\delta^{'}}\left(i\omega\right)}\\&\times\left\{\left[\Gamma^{\Lambda}_{\alpha^{'}\gamma^{'}\alpha\delta}\Gamma^{\Lambda}_{\delta^{'}\beta^{'}\gamma\beta}+(\delta\leftrightarrows \gamma,\delta^{'}\leftrightarrows \gamma^{'})\right]\right.\nonumber\\&-
\left.\left[\Gamma^{\Lambda}_{\beta^{'}\gamma^{'}\alpha\delta}\Gamma^{\Lambda}_{\delta^{'}\alpha^{'}\gamma\beta}+(\delta\leftrightarrows \gamma,\delta^{'}\leftrightarrows \gamma^{'})\right]\right\},\nonumber
%\\&-\sum_{k,k^{'}}\mathcal{S}^{\Lambda}_{k,k^{'}}\Gamma^{\Lambda}_{k^{'}_{1},k^{'}_{2},k^{'};k_{1},k_{2},k},
\label{Gamma}
\end{align}
where ${\mathcal{G}}^{\Lambda}_{\sigma}(i \omega)=\left[\left[\mathcal{G}^{\Lambda}_{0,\sigma}(i\omega)\right]^{-1}-\Sigma^{\Lambda}_{\sigma}\right]^{-1}$, the Greek multi-indices collect the dot and spin indexes and repeated indices imply summation. The $\mathcal{S}^{\Lambda}$ denotes the single-scale propagator (from now, we omit dots indexes for brevity)
\begin{equation}
\mathcal{S}^{\Lambda}_{\sigma}=\mathcal{G}^{\Lambda}_{\sigma}\partial_{\Lambda}\left(\mathcal{G}^{\Lambda}_{0,\sigma}\right)^{-1}\mathcal{G}^{\Lambda}_{\sigma}.
\end{equation}

As in Refs. \cite{Andergassen,Karrasch} we neglect frequency dependence of the self-energy and two-particle vertices, which was previously shown to describe very accurately conductivity of the single-impurity Anderson model in both, weak- and strong coupling regimes; taking into account frequency dependence is not expected to improve results, because of the neglect of the three-particle vertices, see Refs. \cite{Pruschke,Karrasch_thesis}.  \par

The interacting part of the energy energy can be obtained from the flow equation \cite{Metzner,Karrasch_thesis}
\begin{equation}
\partial_{\Lambda}E_{\rm int}^{\Lambda}=\int \dfrac{d\omega} {2\pi} e^{i\omega 0^{+}}\sum_{\alpha}{\left[\left(\mathcal{G}^{\Lambda}_{0}-\mathcal{G}^{\Lambda}\right)\partial_{\Lambda}\left[\mathcal{G}^{\Lambda}_{0}\right]^{-1}\right]_{\alpha\alpha}}.
\end{equation}

The corresponding conductance of a double quantum dot system is given by the relation (see, e.g., Ref.~\cite{Karrasch})
\begin{equation}
 G=2G_0\sum_{\sigma}{|\sum_{i,j}(\Gamma_{i}^{L}\Gamma_{j}^{R})^{1/2}\mathcal{G}^{\Lambda\rightarrow 0}_{ji,\sigma}(0)|^2},
\label{G(Vg)} 
\end{equation}
where $G_{0}=\dfrac{2e^{2}}{h}$ is the conductance quantum~\cite{Oguri_5}.

\subsection{fRG approaches without counterterm}
\begin{figure}[t] 
\centering
\vspace{-0.1cm}
\includegraphics[width=0.9\linewidth]{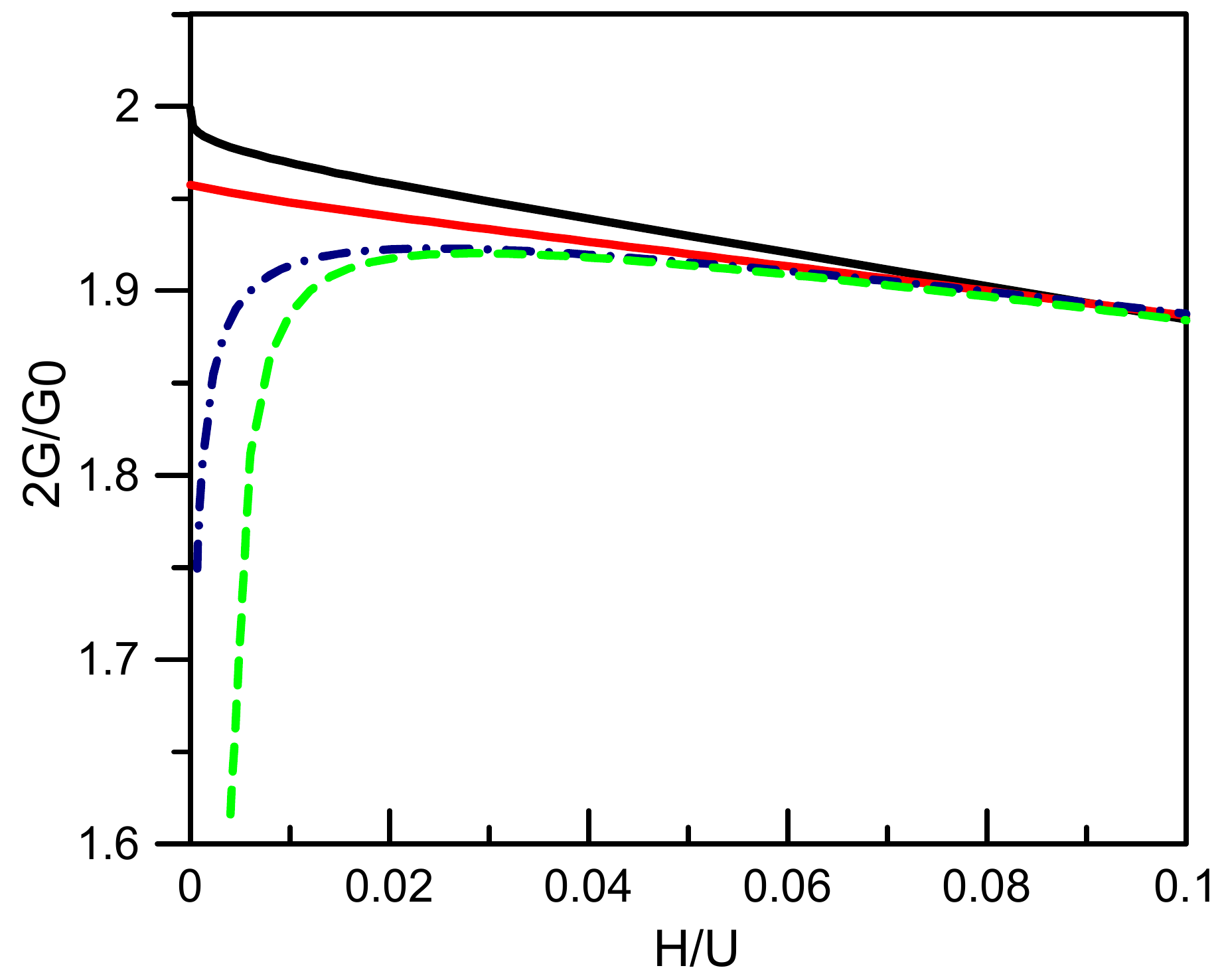}
\vspace{-0.1cm}
\caption{(Color online). Linear conductance $2G/G_{0}$ ($G_0=2e^2/h$) at $T=0$ as a function of magnetic field $H$ for the symmetric case $\Gamma^\alpha_i=U/4$ and $V_{g}=0$. Dashed green and dashed-dotted blue lines correspond to  fRG approximation in the sharp-cutoff and Litim-type cutoff scheme, respectively. Solid black upper line: NRG calculation. Solid red line: fRG approach with the counterterm $\chi_{1}^{\Lambda}$, Eq. (\ref{chi}) with $\tilde{H}=0.1U$ and $\Lambda_{c}=0.05U$.}
\label{G_1_H}
\end{figure}

The standard fRG scheme to study a correlated quantum dots uses sharp cutoff in frequency space \cite{Andergassen,Karrasch,Meden_1,Meden_2} 
\begin{equation}
\mathcal{G}_{0}^\Lambda=\Theta(|\omega|-\Lambda)\mathcal{G}_{0},
\end{equation}
where Heaviside theta function $\Theta$ cuts out infrared modes with Matsubara frequency $|\omega|<\Lambda$. 
Fig.~\ref{G_1_H} shows the zero temperature linear conductance $G$ 
as a function of a magnetic field $H$ at the half-filling $(V_{g}=0)$ in the particular case of full coupling symmetry $t^\alpha_i=t$.
One can see that there is a clear qualitative difference between the magnetic field $H$ dependence of the conductance $G(H)$ at the half-filling $(V_{g}=0)$ within the NRG approach, applied following Refs. \cite{Bulla,NRG,Oguri_1}, and the fRG approach based on the sharp cutoff scheme. The conductance obtained within the latter approach first smoothly increases with decreasing magnetic field and then
suddenly drops when the magnetic field becomes sufficiently small, in contrast to the NRG result.
This drop of the conductance originates from the unphysical behavior of the vertex functions in the fRG flow, 
namely, with decreasing of the magnetic field they first converge to finite, but unphysically large values, and, for even smaller fields, diverge at $\Lambda\rightarrow 0$.  
\par

The results of the approach with sharp cutoff can be somewhat improved using Litim-type \cite{Litim} $\Lambda$--dependence of the bare propagator \cite{Conf}
\begin{multline}
\left[\mathcal{\tilde{G}}^{\Lambda}_{0,\sigma}\right]^{-1}=\left[\mathcal{G}_{0,\sigma}\right]^{-1}+iI\left(\Lambda-|\omega|\right)\\\times\Theta\left(\Lambda-|\omega|\right)\sgn(\omega),
\label{Litim}
\end{multline}
where $I$ is the 2$\times$2 identity matrix, with respect to the indexes of the dots. Using this cutoff, we find that over the whole range of studied magnetic fields the conductance is larger than that obtained in the fRG approach with the sharp momentum cutoff (see Fig. ~\ref{G_1_H}), which leads to some improvement of agreement with NRG data for not too low magnetic fields and
allows us to 
continue the renormalization group flow toward a weaker magnetic fields.

Despite some improvement of the results, the smooth cutoff itself does not allow to 
describe correctly 
conductivity  
at small magnetic fields.  
Furthermore, we found that at low magnetic field these fRG schemes break down not only for the half--filled symmetric case ($V_{g}=0, t^\alpha_i=t$), as considered above, but also in both, symmetric and asymmetric cases, when $\left|V_{g}\right|<V_g^c$, where $V_g^{c}$ is a critical value of gate voltage, which is discussed below. We consider this situation as rather general, and related to the formation of the SFL state in zero magnetic field, that according to the phase diagram obtained in Ref.~\cite{Zitko_2} occurs near half filling.

\subsection{fRG apprach with the counterterm}
To describe the SFL state we decrease the magnetic field continuously during the flow, similarly to a counterterm extension of the fRG \cite{Gersch_1,Gersch_2}.
To this end, we introduce additional term $\sigma{\chi^{\Lambda}}/{2}$ in the
propagator of the quantum dots 
\begin{equation}
\mathcal{G}^{\Lambda}_{0,\sigma}=\left\{\left[\mathcal{\tilde{G}}^{\Lambda}_{0,\sigma}\right]^{-1}+ \dfrac{\sigma}{2} I \chi^{\Lambda} \right\}^{-1}, 
\label{CT}
\end{equation}
corresponding to additional $\Lambda$--dependent 
external magnetic field 
$\chi^{\Lambda}$, which is switched off in the end of the flow, 
if we set $\chi^{\Lambda\rightarrow 0}=0$, and plays a role of infrared regulator in the bosonic spin sector. 
 In this study, we use two different counterterms with linear and exponential cutoff dependence 
\begin{align}
%\chi^{\Lambda}_{1}&=\begin{cases}
%\tilde{H},&\text{if $\Lambda>\Lambda^{c}$;}\\
%\tilde{H}\left(\Lambda / \Lambda^{c}\right),&%\text{if $\Lambda<\Lambda^{c}$},
%\end{cases}\quad\text{or}\notag\\
\chi^{\Lambda}_{1}&=
\tilde{H} \min (1,\Lambda/\Lambda_{c}  ),\notag\\
\quad\chi^{\Lambda}_{2}&=\dfrac{\tilde{H}}{1+\exp{\left[\left(\Lambda_{c}-\Lambda\right)/\Lambda_0\right]}}, 
\label{chi}
\end{align}
that allow to begin the fRG flow with state at the field  $H_{\rm ini}=H+\chi^{\Lambda\rightarrow\infty}=H+\tilde{H}$, while
at the end of the fRG flow one obtains renormalized vertices describing the system in the physical field $H_{\rm fin}=H$; $\Lambda_{c}$, $\tilde{H}$ and $\Lambda_0$ 
are independent parameters, ($\Lambda_c\gg \Lambda_0$),
%; their choice is discussed in the Appendix B. \par 
which determine the scale and sharpness of switching off the $\widetilde{H}$. 

In the present study, the parameter $\Lambda_{0}$ was fixed to be approximately equal to the fRG scale, where the splitting occurs, but can be changed in a rather broad range. The value of the parameter $\Lambda_{c}$ is chosen according to the counterterm field $\tilde{H}$: it should provide a slow switching off the counterterm, when the additional magnetic field is chosen to be small and at the same time the counterterm should not be switched off too slowly for large values of $\tilde{H}$, otherwise significant errors might occur. In other respects these parameters are arbitrary.\par

One can see from Fig.~\ref{G_1_H} that use of the counterterm technique eliminates the unphysical behaviour of the conductance at low magnetic fields. 
In the presence of counterterm, $\Sigma^{\Lambda\rightarrow 0}_{\sigma}$ converges to a finite value,  which entails the nonzero value of the conductance
and significantly improves agreement with the NRG data. 
Although the calculated conductance deviates slightly from the corresponding NRG results, the conductance per spin almost reaches the unitary limit value at zero magnetic field, $\left.G_{\sigma}(H=0)\right|_{V_{g}=0}\approx 0.98e^2/h$.\par

\begin{figure}[htb]
\centering
\includegraphics[width=0.8\linewidth]{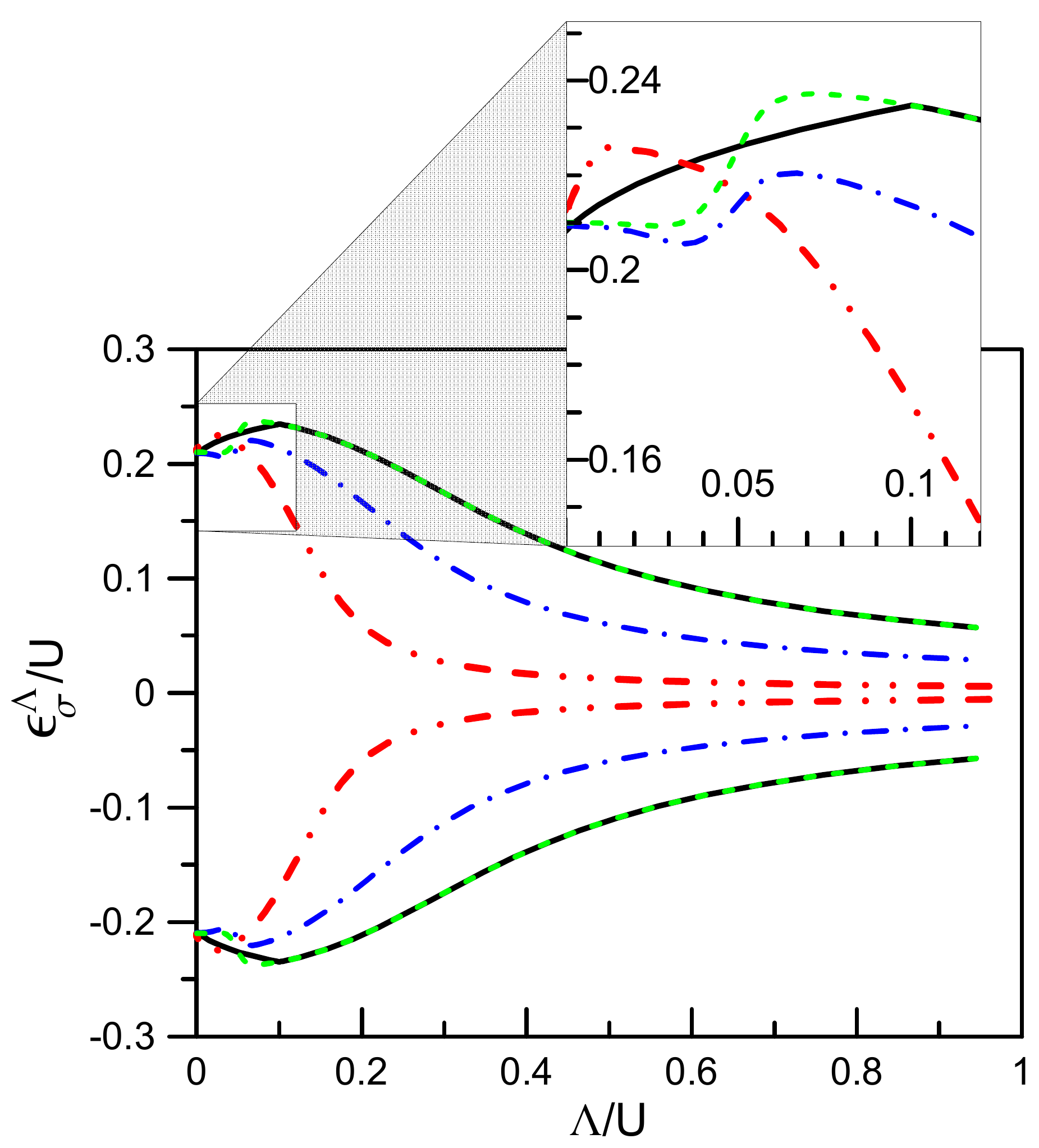}
\caption{(Color online).  Renormalization of the energy levels $\epsilon^{\Lambda}_{\sigma}=\epsilon^{\Lambda}_{1(2),\sigma}$ with spin up $\sigma=\uparrow$
(lower curves) and down $\sigma=\downarrow$ (upper curves) in the fRG approximation with different counterterms for $\Gamma^\alpha_i=U/4$ (note that in this case $\epsilon^{\Lambda}_{1,\sigma}=\epsilon^{\Lambda}_{2,\sigma}$), and $H=V_{g}=0$. Solid (black) line and dashed-dotted-dotted (red) lines: the linear counterterm $(\Lambda_{c}=\tilde{H}/2)$ with $\tilde{H}=0.2U$ and $\tilde{H}=0.02U$, respectively. Dashed (green) and dashed-dotted (blue) lines: exponential counterterm $(\Lambda_{c}=10\Lambda_0=0.05U)$ with $\tilde{H}=0.2U$ and $\tilde{H}=0.1U$, respectively.}
\label{fRG_flows}
\end{figure}

We have verified that the proposed fRG scheme is stable with respect to a choice of the actual form of the counterterm, as well as parameters $\Lambda_{c}$, $\tilde{H}$, and $\Lambda_0$. In particular, in Fig.~\ref{fRG_flows} we plot the dependence of the effective energy levels of the quantum dots $\epsilon^{\Lambda}_{j,\sigma}=\epsilon^{0}_{\sigma}+\Sigma^{\Lambda}_{jj,\sigma}$ on the cutoff parameter $\Lambda$ in different schemes. One can see, that although the flow of the levels is scheme dependent, the final result does not depend on the scheme and initial magnetic field; the latter can be varied in a rather broad range.\par 

\section{Spin splitting, local moments in the SFL phase, and the conductance}
\subsection{Symmetric coupling to the leads}\label{Symm_coupl}

\begin{figure}[t]
\hspace{0.5cm}
\center{\includegraphics[width=0.85\linewidth]{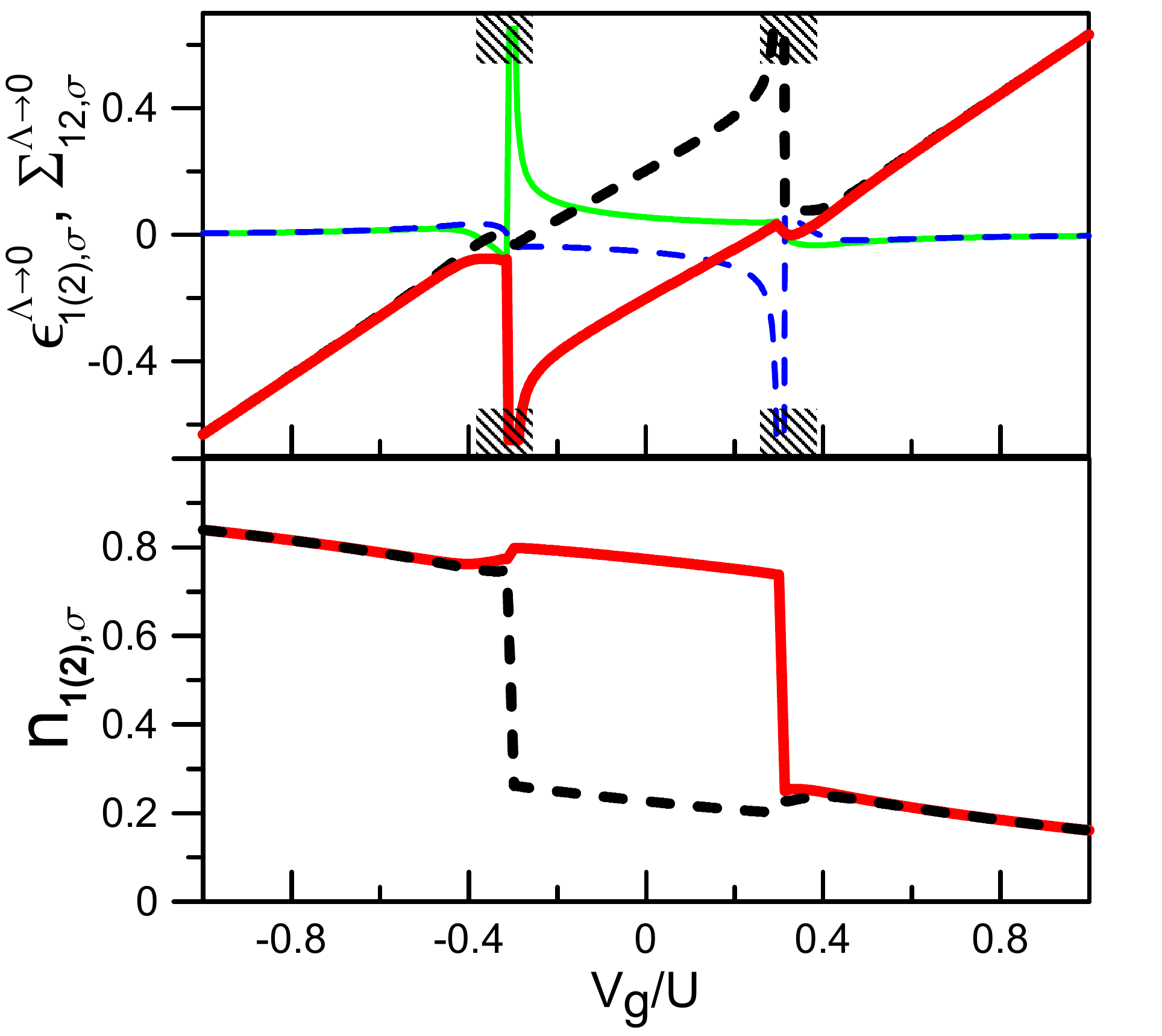}}
\caption{Upper panel: Effective energy levels $\epsilon^{\Lambda\rightarrow 0}_{1(2),\sigma}$ (thick solid (red) and dashed (black) line for $\sigma=\uparrow,\downarrow$, respectively) and off-diagonal self-energies (effective hopping between the levels) $\Sigma^{\Lambda\rightarrow 0}_{12,\sigma}$ 
(thin solid (green) and dashed (blue) line for $\sigma=\uparrow,\downarrow$) in units of $U$ as a function of the gate voltage for $\Gamma^\alpha_i=U/4$ and $H\rightarrow 0$. Lower panel: the average occupation numbers $\langle n_{1(2),\sigma}\rangle$ (solid (red) line and dashed (black) line for $\sigma=\uparrow,\downarrow$) as a function of the gate voltage for the same parameters. The calculations were performed within the fRG approach with the linear counterterm $(\tilde{H}/U=0.1, \Lambda_{c}/U=0.05$).
}
\label{FigS}
\end{figure}

The results of the calculation of spin-dependent self-energies $\Sigma^{\Lambda\rightarrow 0}_{ij,\sigma}$ (in the following the renormalized energy levels $\epsilon^{\Lambda\rightarrow 0}_{j,\sigma}$ are used as a measure of the diagonal diagonal self-energies $\Sigma^{\Lambda\rightarrow 0}_{ii,\sigma}$) and occupation numbers $
\langle n_{j,\sigma}\rangle=%\dfrac{1}{2\pi}\int{d\omega e^{iw0^{+}}{\mathcal{G}}_{jj;\sigma}^{\Lambda\rightarrow0}\left(iw\right)}\\
\int{\dfrac{d\omega}{2\pi} e^{i\omega 0^{+}}\left[\left[{\mathcal{G}}_{0,\sigma}^{\Lambda=0}\left(i\omega\right)\right]^{-1}-\Sigma_{\sigma}^{\Lambda\rightarrow0}\right]^{-1}_{jj}}$ at $T=0$ are shown in Fig.~\ref{FigS}. One can see that the self-energies and occupation numbers 
are strongly split (with respect to spin projection) in some range $|V_g|<V_g^{c}$ near the half-filling $V_{g}=0$ even when $H \rightarrow 0$. As we discuss below, this reflects the formation of local moment in this range of gate voltages. There is small splitting of self-energies also in the vicinity of the transition point (in the shaded area) in the paramagnetic phase $|V_{g}|>V^{c}_{g}$, which is an artifact of the present approach.
We also note, that in the vicinity of the transition point, the proposed fRG approach may overestimate the renormalization of the vertices and for this reason may become less accurate for $V_{g}\approx V_{g}^{c}$  (in the shaded areas in the Fig.~\ref{FigS}a). However, we have found that even for these gate voltages all the observables are described correctly in comparison with the NRG calculations.\par

The local moment appears due to the specific charge redistribution, which in the symmetric case $t^\alpha_i=t$ is related to the even and odd orbitals $d_{e(o),\sigma}$,
yielding one electron in the odd orbital, disconnected from leads, and changing the character of the spin-spin correlations \cite{Zitko_2}. 
 \begin{figure}[htb]
\center{\includegraphics[width=0.8\linewidth]{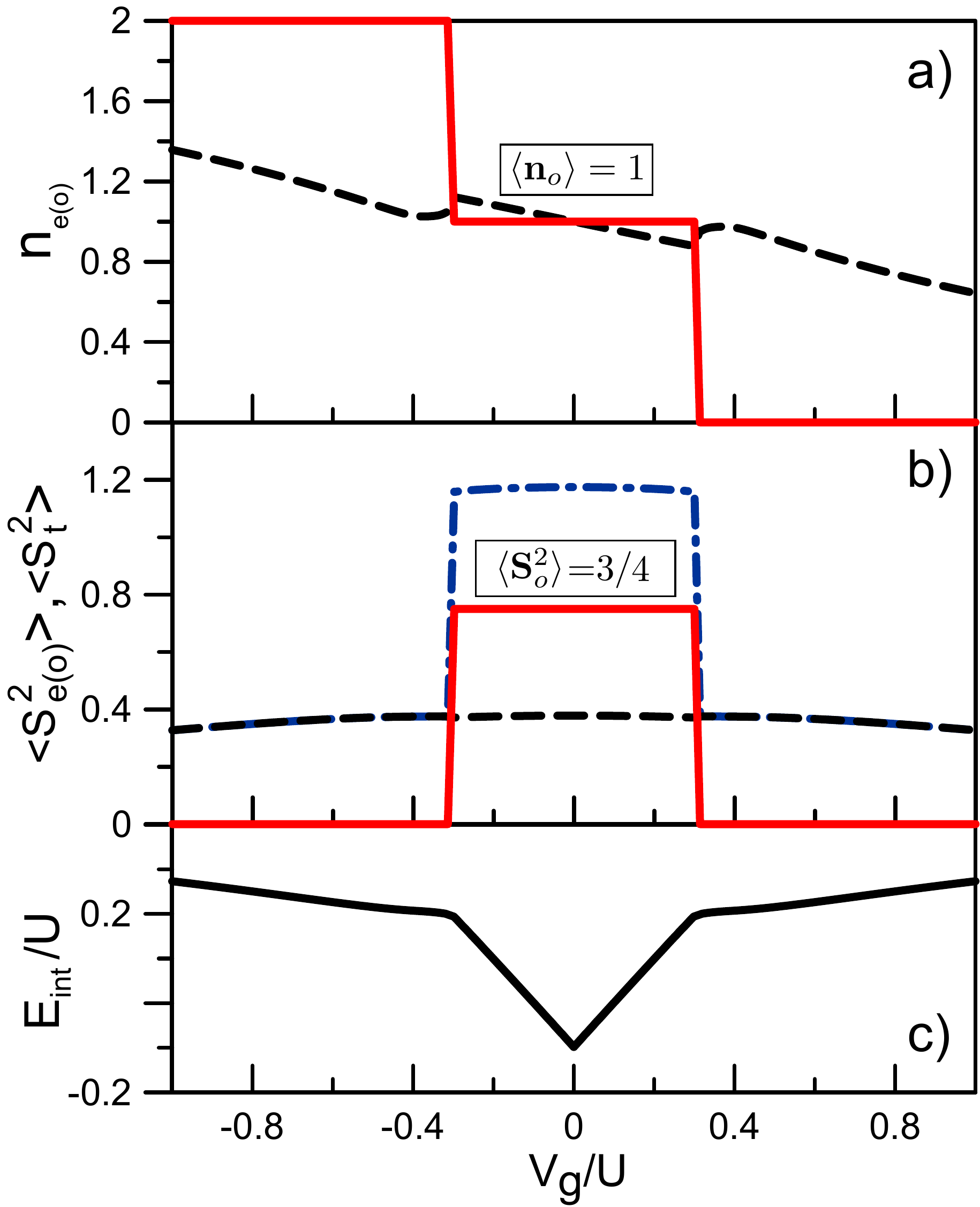}}  
\caption{(Color online). a) and b) The occupation numbers $\langle n_{e(0)}\rangle$ and the average square of magnetic moment $\langle\mathbf{S}_{e(0)}^{2}\rangle$ in the even (dashed (black) lines) and odd (solid (red) lines) orbitals, as well as the average of the square of the total spin $\langle\mathbf{S}_{t}^{2}\rangle=\langle\left(\mathbf{S_{1}+S_{2}}\right)^{2}\rangle$ (dashed-dotted (blue) line) as a function of the gate voltage for $\Gamma^\alpha_i=U/4$;
(c) The interaction energy $E_{\rm int}$ of the double quantum dot system as a function of ${V_{g}/U}$.}
\label{Occ_levels}
\end{figure}
To show this explicitly, we plot the total occupation numbers $\langle n_{e(o)}\rangle=\sum_{\sigma}{\langle n_{e(o),\sigma}\rangle}=\sum_{\sigma}\langle d^{\dagger}_{e(o),\sigma}d_{e(o),\sigma}\rangle $ in Fig. \ref{Occ_levels}(a). In agreement with Ref.~\cite{Zitko_2} in finite range of gate voltages $|V_g|<V_g^c$  
the odd orbital  is occupied by one electron $\langle n_{o}\rangle=%\sum_{\sigma}{\langle n_{o,\sigma\rangle}=
1$,  forming a $S = 1/2$ local moment, which is aligned along the direction of the magnetic field.  
 The occupation numbers  $\langle n_{e,\sigma}\rangle$ for the even orbital behave 
 smoothly 
 due to the level broadening caused by the hybridization of the even orbital with leads.
In Fig. \ref{Occ_levels}(b) we also plot 
the gate voltage dependence of the square of the spin in the even(odd) orbitals $\langle\mathbf{S}^{2}_{e(o)}\rangle=(1/4)\sum_{\sigma,\sigma^{'}}\langle (d^{\dagger}_{e(o),\sigma}{\bm{\sigma}}d_{e(o),\sigma^{'}})^2\rangle$ (${\bm{\sigma}}$ are the Pauli matrices). 
The square of the spin on the odd orbital, $\langle\mathbf{S}^{2}_{o}\rangle$  is equal to $S(S+1)=3/4$ in the region $\left|V_{g}\right|<V_{g}^{c}$ and becomes zero for the $\left|V_{g}\right|>V_{g}^{c}$. In contrast, the square of the spin in the even orbital behaves smoothly and does not shows a significant change at the $V_{g}=\pm V_{g}^{c}$. The average of the square of the total spin $\langle\mathbf{S}_{t}^{2}\rangle$ (see Fig. \ref{Occ_levels}(b)) takes an almost constant value, $\langle\mathbf{S}_{t}^{2}\rangle\approx 1.2$, for $\left|V_{g}\right|<V_{g}^{c}$  and has a weakly pronounced maximum at $V_{g}=0$.  In this region of gate voltage $\langle\mathbf{S}_{t}^{2}\rangle>\langle\mathbf{S}^{2}_{e}\rangle+\langle\mathbf{S}^{2}_{o}\rangle=\langle\mathbf{S}^{2}_{1}\rangle+\langle\mathbf{S}^{2}_{2}\rangle$, which indicates the presence of ferromagnetic correlations. 
%With increasing $\left|V_{g}\right|$ the total square of the spin slightly decreases and at the transition point $V_{g}=V_{g}^{c}$ shows discontinuity, accompanied by a sharp decrease of the amplitude of $\langle\mathbf{S}_{t}^{2}\rangle$. \par

The dependence of the interaction energy $E_{\rm int}=E-E_0$, i.e. a difference between the energy $E$ of interacting and $E_0$ of non-interacting systems, on the gate voltage (shown in Fig.~\ref{Occ_levels}(c))  indicates that the jumps of occupation numbers correspond to the cusp of the $E(V_g)$ dependence, confirming first-order phase transition at $|V_g|=V_g^c$.

In order to understand the mechanism that leads to the existence of spin splitting of energy levels in the SFL state,
we rewrite the Hamiltonian (\ref{H}) for the symmetric case $t^\alpha_i=t$ in terms of even and odd orbitals (up to a constant contribution) as
 \begin{align}  &\mathcal{H_{\rm dot}}=\sum_{k,\sigma}\left(\epsilon^{0}_{\sigma}-\dfrac{U}{2}\right)n_{k,\sigma}-U\vec{\mathbf{S}}_{e}\vec{\mathbf{S}}_{o}+\dfrac{U}{2}\left[\vphantom{\frac{1}{2}}n_{e,\uparrow}n_{e,\downarrow}\right.\notag\\&+\left.n_{o,\uparrow}n_{o,\downarrow}+\dfrac{1}{2}n_{o}n_{e}+ \left(d^{\dagger}_{e,\uparrow}d_{o,\uparrow}d^{\dagger}_{e,\downarrow}d_{o,\downarrow}+\text{H.c.}\right)\right].%+\dfrac{U}{2}.
% \notag
\label{H_e_o}
  \end{align}   
The Hamiltonian (\ref{H_e_o}) has a form of the two-orbital Anderson model with intra-orbital, inter-orbital, Hund exchange interaction, and pair electron hopping equal to $U/2$.
In the SFL (local moment) phase even for the infinitesimally small magnetic field, the Hund exchange leads to strong splitting of itinerant  electronic states 
$$
-U\vec{\mathbf{S}}_{e}\vec{\mathbf{S}}_{o}\sim -\dfrac{U}{4}\sum_{\sigma}{\sigma n_{e,\sigma}}.
%=-\sum_{\sigma}{\sigma\dfrac{H_{ex}}{2}n_{e\sigma}}.
$$
As a result the Zeeman energy splitting of even states becomes $\Delta_{\uparrow\downarrow}\sim %H_{ex}+H\sim 
U/2$,
instead of $\Delta_{\uparrow\downarrow}=H$ in the absence of interaction $U$, and 
therefore it is strongly enhanced. Due to half filling of the odd orbital, the pair hopping term is not active in the SFL phase.
\par 
The results
showing spin splitting in SFL phase refer to $H \rightarrow 0$ limit of Green functions at $T=0$. Presence of the local moment makes however non-commutative limits $T\rightarrow 0$ and $H \rightarrow 0$. 
In the opposite limit $H=0$, $T\rightarrow 0$ our NRG calculations confirm the validity of 
Logan et al. result \cite{Logan}, expressing the Green function through the arithmetic average of spin-split Green functions $\mathcal{G}(H=0)=(\mathcal{G}_\uparrow+\mathcal{G}_\downarrow)/2$, calculated at small finite magnetic field (which can be therefore extracted from fRG calculation).\par
\begin{figure}[t!]
\center{\includegraphics[width=0.85\linewidth]{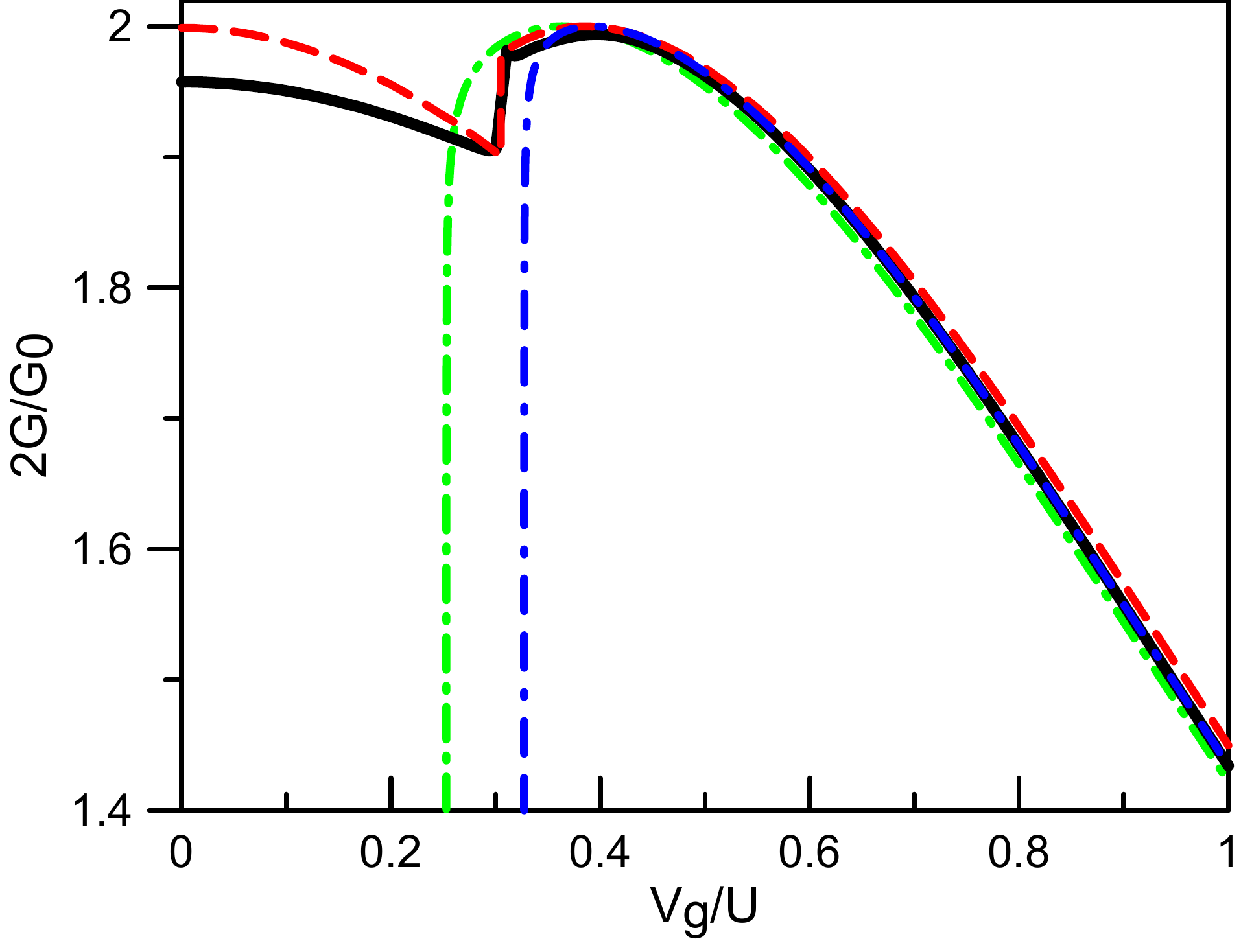}}\\
\center{\includegraphics[width=0.85\linewidth]{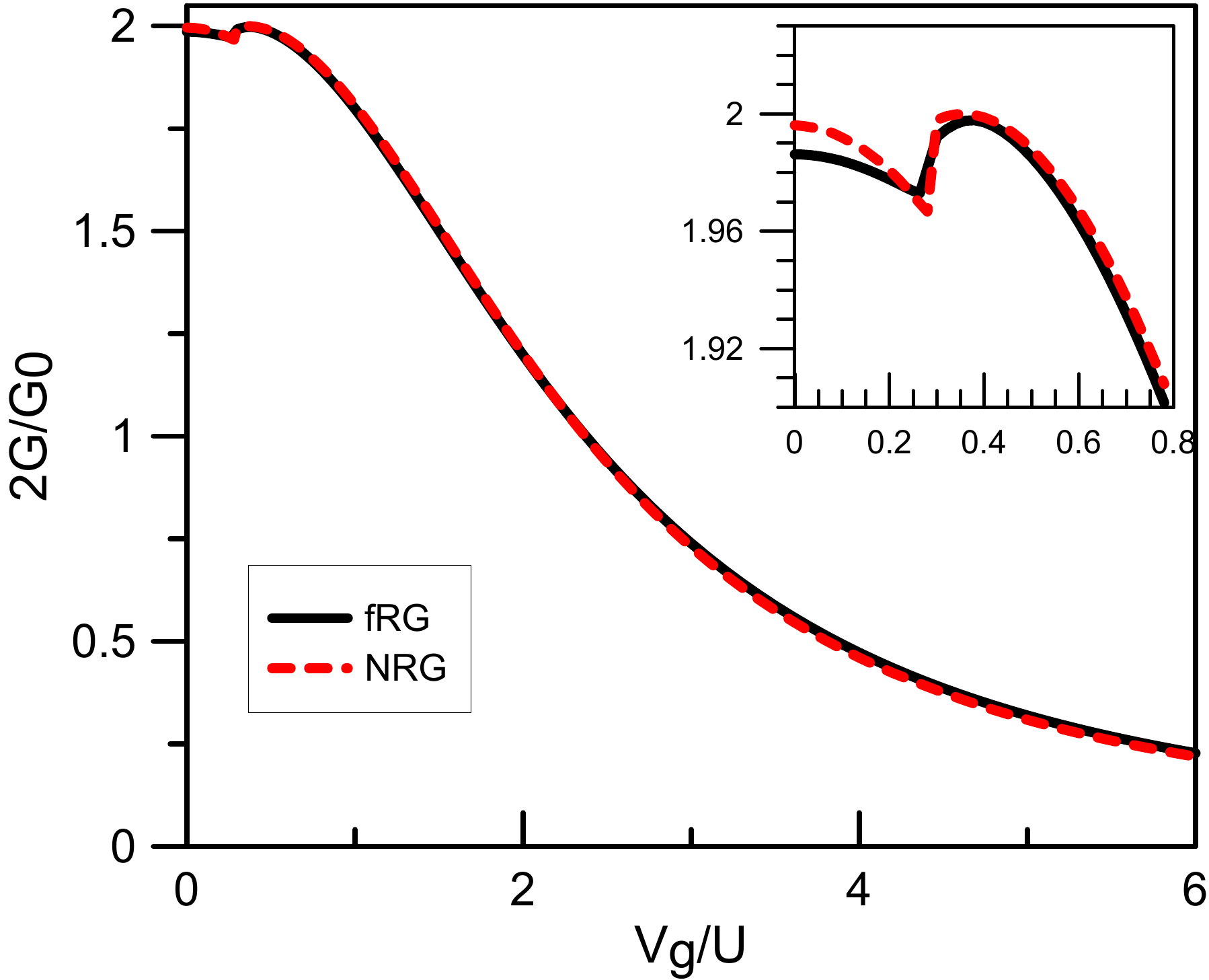}}\par
\center{\includegraphics[width=0.85\linewidth]{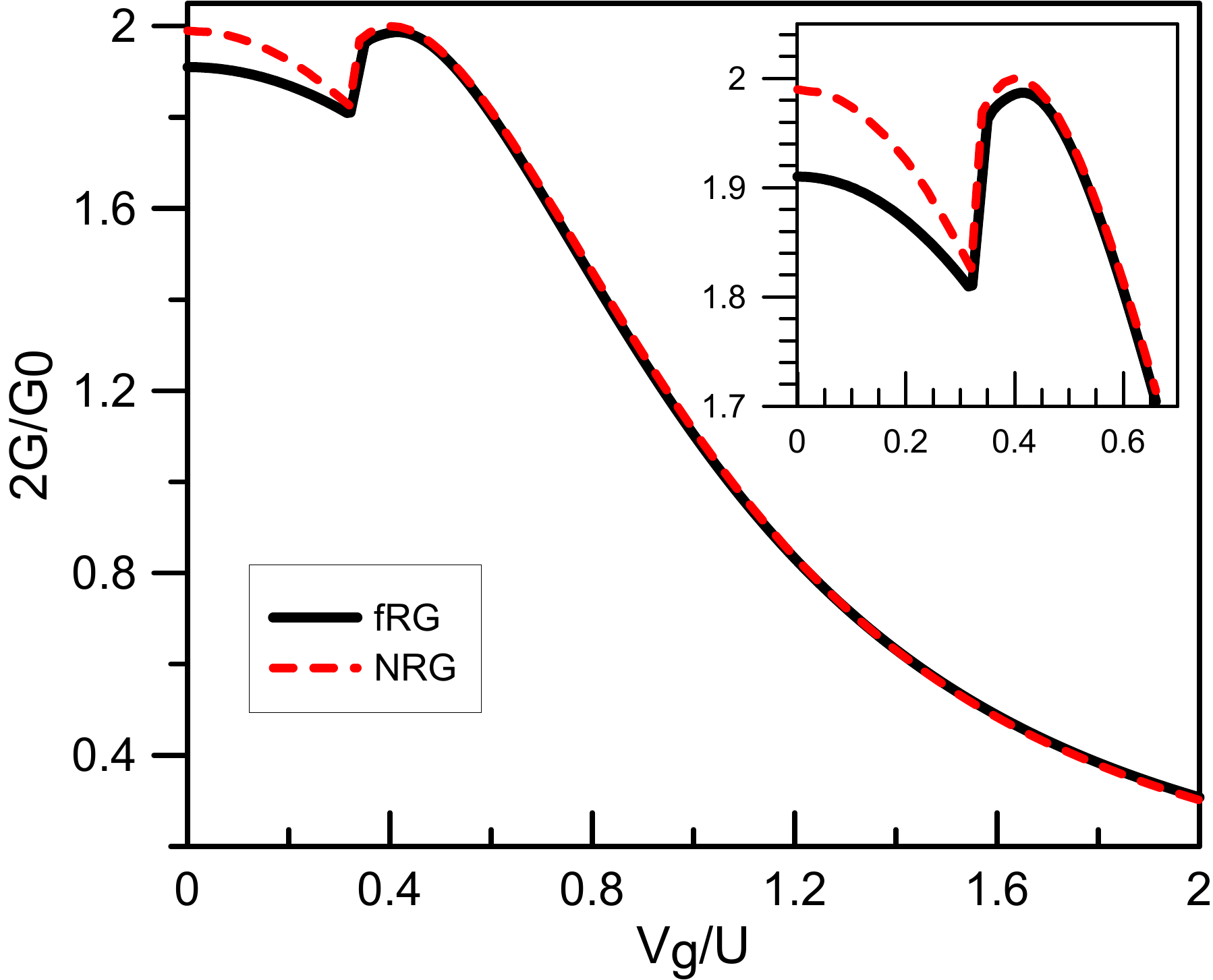}}
\caption{(Color online). The dependence of linear conductance $G$ on gate voltage ${V_{g}}$ for $\Gamma^\alpha_i=U/4$ (upper plot), $\Gamma^\alpha_i=U/2$ (middle plot) and $\Gamma^\alpha_i=U/6$ (lower plot), $H\rightarrow 0$ (at temperature $T=0$) within the fRG approach with the linear counterterm $(\tilde{H}/U=0.1, \Lambda_{c}/U=0.05$, solid black line) and NRG calculation (dashed red line). Dashed-dotted green and dashed-dashed dotted blue lines on upper plot are fRG approaches in the sharp-cutoff and Litim-type cutoff scheme without counterterm. 
}
\label{FigG}
\end{figure}

 
%\section{Conductance} 
The resulting dependence of the conductance $G(V_{g})$, obtained at $T=0$ and $H\rightarrow 0$ is represented in Fig.~\ref{FigG}.   
We can see that while the conductance within the both sharp and smooth cutoffs drops to zero in the SFL state,
the fRG approach with the cutoff procedure (\ref{CT}) leads to a finite conductance in the whole range $|V_g|<V_g^c$. The conductance in fRG almost reaches unitary value $G_{0}=2e^2/h$ at $V_{g}=0$ and in the symmetric case $t_i^\alpha=t$ shows discontinuity 
at the gate voltage $V^c_{g}$, corresponding to a quantum
phase transition  from SFL to a regular FL ground state, which agrees well with our NRG results and previous NRG analysis 
for strong on-site Coulomb repulsion \cite{Zitko_1}. It can be seen that the conductance in the singular Fermi liquid regime is well described by the proposed approach; as expected, the agreement with NRG becomes better as $U/\Gamma$ is decreased. At the same time, for $U/\Gamma_i^\alpha=6$, the disagreement with NRG is still not too large close to half-filling. Furthermore, we have verified that this holds for $U/\Gamma_i^{\alpha}\lesssim 10$.

The conductance in the opposite limit $T\rightarrow0$ at $H=0$ may differ from the above result due to additional phase shift \cite{Logan}. 
Neglecting vertex corrections, the conductance in this limit is given by $G=4G_0{|\sum_{i,j}(\Gamma_{i}^{L}\Gamma_{j}^{R})^{1/2}\mathcal{G}^{\Lambda\rightarrow 0}_{ji}(H=0)|^2}$, which yields somewhat smaller value, than at $T=0$, $H\rightarrow 0$.\par

%In order to provide a thorough comparison between the functional renormalization group and numerical renormalization group methods, we also plot the linear conductance for stronger  $(\Gamma_i^\alpha=U/6)$ and weaker $(\Gamma_i^\alpha=U/2)$ Coulomb interaction in Fig. \ref{FigG}.  

\begin{figure}[t]
\center{\includegraphics[width=1\linewidth]{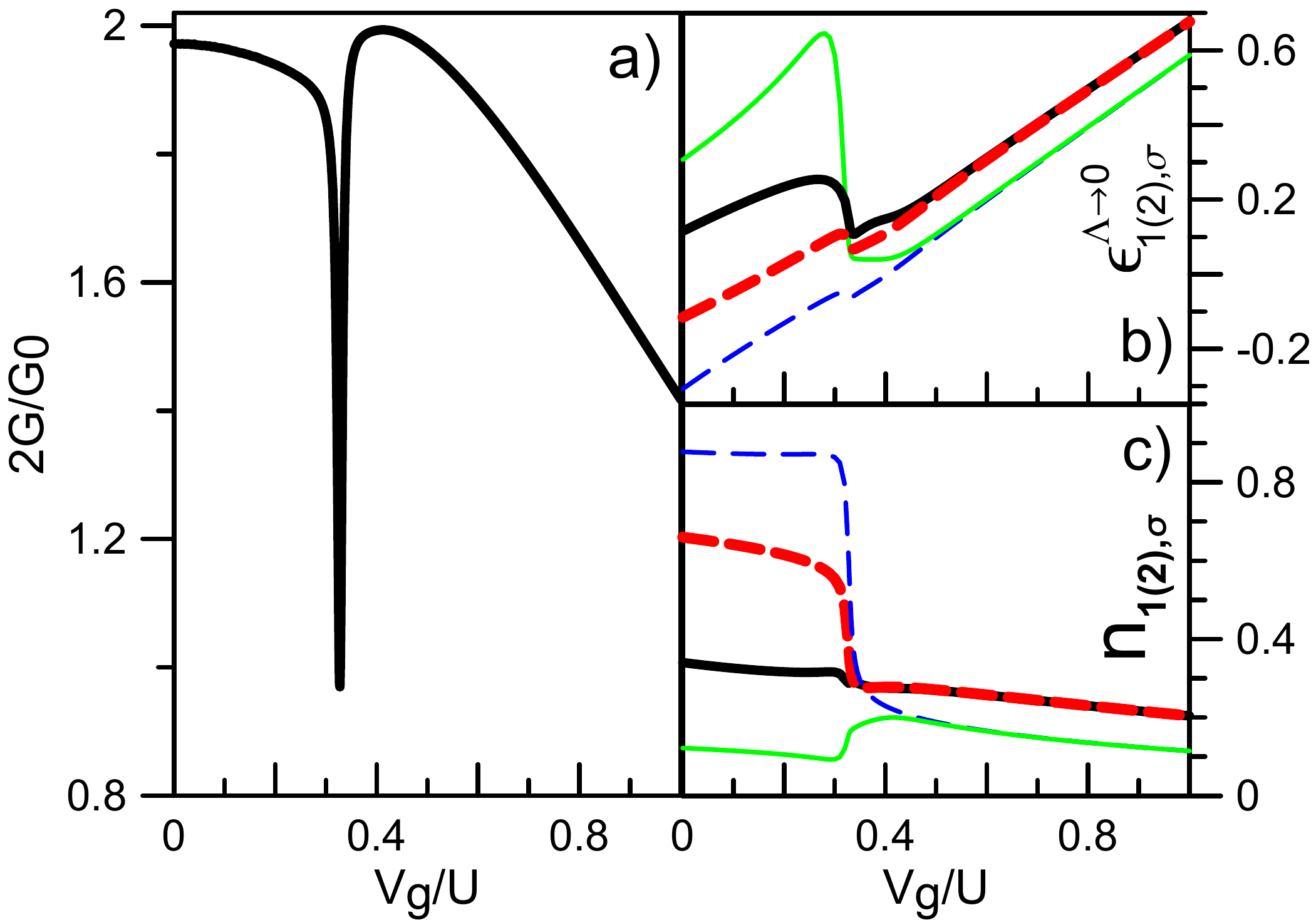}}
\caption{(Color online). The  conductance (a), effective energy levels (b) $\epsilon^{\Lambda\rightarrow 0}_{1,\sigma}$ (thick dashed (red) line for $\sigma=\uparrow$ and thick solid (black) line for $\sigma=\downarrow$) and $\epsilon^{\Lambda\rightarrow 0}_{2,\sigma}$ (thin dashed (blue) line for $\sigma=\uparrow$ and thin solid (green) line for $\sigma=\downarrow$), as well as the spin--resolved average occupation numbers (c) of each quantum dot $\langle n_{1(2),\sigma}\rangle$ (with the same notation of the 
curves as on (b)) in the asymmetric case with $\Gamma_{1}^{L}=\Gamma_{1}^{R}=U/3$ and $\Gamma_{2}^{L}=\Gamma_{2}^{R}=U/6$ obtained by the fRG approach with the linear counterterm $(\tilde{H}/U=0.1, \Lambda_{c}/U=0.05$).
}
\label{FigG1}
\end{figure} 

\subsection{Asymmetric coupling to the leads}

In order to consider influence of asymmetric coupling to the leads on the SFL phase and behaviour of the conductance we also carried out the fRG calculation for the cases, when the coupling part of the double quantum dot Hamiltonian have the more general form with different hopping parameters $t_i^\alpha$.
It turns out that for rather different sets of the hopping parameters we face again with the problem of the unphysical behaviour of the vertex functions in the standard fRG schemes at low magnetic field; as for the isotropic case the fRG approach with the counterterm fixes this problem and physical behavior near the half--filling can be described.

As an example, in Fig.~\ref{FigG1} we plot the conductance (a), effective energy levels (b) and occupation numbers (c) of the parallel quantum dot system with up-down coupling asymmetry  $\Gamma_{1}^{L}=\Gamma_{1}^{R}=U/3$ and $\Gamma_{2}^{L}=\Gamma_{2}^{R}=U/6$. From Fig.~\ref{FigG1}b) one can see that as in the perfectly symmetric case there exists a critical gate voltage $V_{g}^{c}$ below which energy levels acquire again finite spin splitting in the presence of infinitesimal magnetic field. As expected, in the case when the quantum dots are asymmetrically connected to leads the value of the splitting become different for both quantum dots. For $|V_{g}| < V_{g}^{c}$ the energy levels of  quantum dot, weakers hybridized with the leads, are strongly split, which leads to a significant difference between the average occupation numbers of up- and down-spin states for this dot (see Fig.~\ref{FigG1}c). In contrast, energy levels of the quantum dot $\epsilon_{1,\sigma}$, stronger hybridized with the leads, do not exhibit such a strong splitting and the corresponding occupation numbers $\langle n_{1,\sigma}\rangle$ are more sensitive to a change of the gate voltage. The overall behavior of the renormalized energy levels and occupation numbers is nevertheless similar to that obtained for the symmetric coupling to the leads, showing that SFL state is not restricted to the symmetric coupling of quantum dots to the leads. The main difference with the symmetric case is that the transition to SFL state is continuous in the presence of the asymmetry. The absence of the  discontinuities in the asymmetric case is a consequence of the interaction induced non-zero hopping between even and odd orbitals. The corresponding transition to SFL phase can be therefore viewed as a (second-order) quantum phase transition.

The conductance of parallel quantum dot system with up-down coupling asymmetry shown in Fig.~\ref{FigG1}a) has also no discontinuities and shows a sharp asymmetric Fano--like resonance at a gate voltage $V_{g}^{c}$, indicating the SFL behavior. The NRG calculations for asymmetric cases become progressively more complicated (especially when all hopping parameters are different), and they are not therefore presented here. Due to rich phase diagram of these asymmetric cases, with several independent hopping parameters and gate voltage, a more detail analysis of these cases will be given in the upcoming publication~\cite{Asymm}.

\section{Conclusion}
In summary, we have proposed the version of the functional renormalization-group approach, which is able to describe interaction-induced local moments in quantum dot systems, which occur due to peculiarities of geometric structure of these systems. We applied the proposed scheme to obtain occupation numbers, square of the spin, and conductance as a function of gate voltage in parallel quantum dots, symmetrically coupled to leads, and found good agreement with NRG calculations. Investigation of the gate voltage dependence of the interaction energy shows confirms first order quantum phase transition at the gate voltage $V_g^c$ where abovementioned quantities show a jump. For parallel quantum dots, differently coupled to the leads, the conductance shows Fano-like resonance. 

The proposed approach can be further used for describing equilibrium and non-equilibrium processes in complex quantum dot systems or organic molecules, which are not accessible for NRG approach. The presence of local moments in some geometries of these systems can be in particular utilized in quantum computation devices. 

{\it Acknowledgements}. The work is performed within the theme ”Electron”
01201463326 of FASO, Russian Federation. 
%, and partially
%supported  within  the  Act  211  Government  of  the  Russian  Federation,  contract  02.A03.21.0006.   
Calculations
were performed on the Uran cluster of Ural branch RAS.
%\section*{Appendix A: The single scale propagator and its integrals in the fRG approach with the counterterms }

% Using the explicit form of the cutoff dependent noninteracting Green function~(\ref{CT}), the single-scale propagator can be written as
%\begin{align}
%\mathcal{S}_{\sigma}^{\Lambda}(i\omega)&=\tilde{\mathcal{S}}_{\sigma}^{\Lambda}(i\omega)+\mathcal{S}_{\sigma}^{\chi}(i\omega)=i\Theta\left(\Lambda-|\omega|\right)\sgn(\omega)\notag\\&\times\left[\mathcal{G}_{\sigma}^{\Lambda}(i\omega)\right]^{2}+\sigma\dfrac{\dot{\chi}^{\Lambda}}{2}\left[\mathcal{G}_{\sigma}^{\Lambda}(i\omega)\right]^{2}.
%\end{align}

%The integrals, appearing at the r.h.s of the fRG equations and containing  $\tilde{\mathcal{S}}_{\sigma}^{\Lambda}(i\omega)$ 
%are analytically computed:
%\begin{equation}
%\int \dfrac{d\omega}{2\pi}\tilde{\mathcal{S}}^{\Lambda}(i\omega)=\dfrac{i\Lambda}{2\pi}\left(\left[\mathcal{G}^{\Lambda}_{+}\right]^{2}-\left[\mathcal{G}^{\Lambda}_{-}\right]^{2}\right),
%\end{equation}
%and
%\begin{multline}
%\int \dfrac{d\omega}{2\pi}\tilde{\mathcal{S}}_{\gamma,\gamma^{'}}^{\Lambda}(i\omega)\mathcal{G}_{\delta,\delta^{'}}^{\Lambda}(\mp i\omega)=\\\dfrac{i\Lambda}{2\pi}\left(\left[\mathcal{G}^{\Lambda}_{+}\right]_{\gamma,\gamma^{'}}^{2}\left[\mathcal{G}^{\Lambda}_{\mp}\right]_{\delta,\delta^{'}}-\left[\mathcal{G}^{\Lambda}_{-}\right]_{\gamma,\gamma^{'}}^{2}\left[\mathcal{G}^{\Lambda}_{\pm}\right]_{\delta,\delta^{'}}\right),
%\end{multline}
%where
%$
%\mathcal{G}^{\Lambda}_{\pm}={\mathcal{G}}^{\Lambda}(\pm i \Lambda)
%$.\par

%\clearpage
\setcounter{equation}{0}
\setcounter{figure}{0}
\renewcommand\theequation{S\arabic{equation}}
\renewcommand\thefigure{S\arabic{figure}}
\makeatletter
\makeatother
\end{document}